# Refined multiscale fuzzy entropy based on standard deviation for biomedical signal analysis

Hamed Azami[1] · Alberto Fernández[2] · Javier Escudero[1]

**Abstract** Multiscale entropy (MSE) has been a prevalent algorithm to quantify the complexity of biomedical time series. Recent developments in the field have tried to alleviate the problem of undefined MSE values for short signals. Moreover, there has been a recent interest in using other statistical moments than the mean, i.e., variance, in the coarse-graining step of the MSE. Building on these trends, here we introduce the so-called refined composite multiscale fuzzy entropy based on the standard deviation ($RCMFE_\sigma$) and mean ($RCMFE_\mu$) to quantify the dynamical properties of spread and mean, respectively, over multiple time scales. We demonstrate the dependency of the $RCMFE_\sigma$ and $RCMFE_\mu$, in comparison with other multiscale approaches, on several straightforward signal processing concepts using a set of synthetic signals. The results evidenced that the $RCMFE_\sigma$ and $RCMFE_\mu$ values are more stable and reliable than the classical multiscale entropy ones. We also inspect the ability of using the standard deviation as well as the mean in the coarse-graining process using magnetoencephalograms in Alzheimer's disease and publicly available electroencephalograms recorded from focal and non-focal areas in epilepsy. Our results indicated that when the $RCMFE_\mu$ cannot distinguish different types of dynamics of a particular time series at some scale factors, the $RCMFE_\sigma$ may do so, and vice versa. The results showed that $RCMFE_\sigma$-based features lead to higher classification accuracies in comparison with the $RCMFE_\mu$-based ones. We also made freely available all the Matlab codes used in this study at http://dx.doi.org/10.7488/ds/1477.

✉ Hamed Azami
  hamed.azami@ed.ac.uk

  Alberto Fernández
  aferlucas@med.ucm.es

  Javier Escudero
  javier.escudero@ed.ac.uk

1 Institute for Digital Communications, School of Engineering, University of Edinburgh, King's Buildings, Edinburgh EH9 3FB, UK
2 Departamento de Psiquiatría y Psicología, Médica, Universidad Complutense de Madrid, Madrid, Spain

## 1 Introduction

An important challenge in signal processing is to quantify the dynamical irregularity of time series [1]. To this end, there are a number of approaches, such as entropies and fractal dimensions. Entropy is an appealing and powerful tool that has been widely used in physiological signal analysis [1, 2]. One of the most popular entropy-based approaches is sample entropy (SampEn), which is relatively robust to noise [2]. Another widely used entropy method is fuzzy entropy (FuzEn) [3]. These two entropy approaches have attracted a great deal of attention recently [4–7]. Although SampEn is slightly faster than FuzEn, the latter is more consistent and less dependent on the data length [3, 7].

The traditional methods to quantifying the complexity of biomedical recordings may fail to account for the multiple time scales inherent in such time series and may yield contradictory and misleading results. For instance, even though the SampEn of white Gaussian noise (WGN) time series is higher than that of $1/f$ noise, showing that WGN is more irregular than $1/f$ noise, the latter has more complex structures than WGN due to the presence of long-range correlations [8, 9].

To address this problem, Costa et al. introduced the multiscale (sample) entropy (MSE), which is based on assessing the entropy of signals at multiple time scales [8]. In the MSE method, the original signal is first divided into non-overlapping segments of length $\tau$, termed the scale factor. Next, the mean of each segment is estimated to derive the coarse-grained signals. Finally, the entropy measure, using SampEn, is calculated for each coarse-grained sequence [8].

The complexity evaluation of time series with MSE is rooted in the concept that complexity is associated with "meaningful structural richness," which may be in contrast with regularity measures defined from classical entropy algorithms [8, 10]. This is because the output of entropy-based metrics grows monotonically with the degree of randomness of the analyzed time series. Therefore, these measures assign the highest entropy values to uncorrelated random signals like white noise, which are highly unpredictable but not structurally "complex," and, at a global level, permit a very simple description. Thus, when applied to biomedical signals, traditional entropy-based methods may lead to misleading outputs. For instance, they assign high entropy values to certain pathologic cardiac rhythms that generate erratic outputs whereas healthy cardiac rhythms that are exquisitely regulated by multiple interacting control mechanisms are given low values of entropy. In this context, the complexity of biomedical signals reflects their ability to adapt and function in an ever-changing environment because physiological signals require to operate across multiple temporal and spatial scales. Thus, substantial attention has been concentrated on defining a quantitative measurement of complexity, i.e., MSE, that vanishes for both deterministic/predictable and uncorrelated random/unpredictable time series [8, 9]. Extensive analyses have shown that abnormal and disease states, which decrease the adaptive capacity of the subject, appear to degrade the multiscale entropy metrics [8, 9]. A recent review about multiscale entropy-based methods can be seen in [11].

Costa and Goldberger have very recently introduced a new MSE approach using the variance, instead of the mean, in the coarse-graining process of MSE. This was named $MSE_{\sigma}^2$ [12]. Note that, in order to discriminate $MSE_{\sigma}^2$ and basic MSE, we will denote the latter as $MSE_{\mu}$. $MSE_{\sigma}^2$ revealed that the dynamics of the volatility (variance) of heartbeat signals obtained from healthy young subjects are highly complex [12].

Nonetheless, since the standard deviation ($\sigma$) has the same dimension as the signal and its mean values ($MSE_{\mu}$), we propose to use $\sigma$ in the coarse-graining process, as an alternative to $MSE_{\mu}$ and $MSE_{\sigma}^2$. Furthermore, one of the most important problems of $MSE_{\mu}$ is that, when applied to short biological signals, the results may be undefined and inaccurate [13, 14]. To alleviate this problem, the refined composite $MSE_{\mu}$ ($RCMSE_{\mu}$) has been recently introduced [13] using the average of the SampEn values of several coarse-grained signals in each scale factor. Although simulation results showed that the $RCMSE_{\mu}$ had better stability for all temporal scales than $MSE_{\mu}$, the problem of undefined values for short signal still exists [13]. We build on these recent developments to combine their advantages, and propose the refined composite multiscale fuzzy entropy (RCMFE) based on $\mu$ and $\sigma$: $RCMFE_{\mu}$ and $RCMFE_{\sigma}$, respectively. We hypothesize that these measures will be more accurate, robust, and stable than previous entropy metrics. Furthermore, we exemplify the behavior of these measures for different kinds of classical signal concepts (e.g., frequency, non-linearity) to demonstrate the dependency of $RCMFE_{\sigma}$ and $RCMFE_{\mu}$ on them. Finally, we illustrate their application to two clinical datasets: focal and non-focal electroencephalograms (EEGs) and resting-state magnetoencephalogram (MEG) activity in Alzheimer's disease (AD).

## 2 Methods

### 2.1 Entropy approaches

#### 2.1.1 Sample entropy

Assume we have a real-valued discrete time series of length $N$: $\mathbf{y} = \{y_1, y_2, ..., y_N\}$. At each time $t$ of $\mathbf{y}$, a vector including the $m$-th subsequent values is constructed as $Y_t^m = \{y_t, y_{t+1}, ..., y_{t+m-2}, y_{t+m-1}\}$ for $t = 1,2,...,N-(m-1)$, where $m$, termed embedding dimension, determines how many samples are contained in each vector. Define the distance between such vectors as the maximum difference of their corresponding scalar components, $d\left[Y_{t_1}^m, Y_{t_2}^m\right] = \max\left\{\left|Y_{t_1+k}^m - Y_{t_2+k}^m\right| : 0 \leq k \leq m-1 \text{ and } t_1 \neq t_2\right\}$. A match happens when the distance $d\left[Y_{t_1}^m, Y_{t_2}^m\right]$ is smaller than a predefined tolerance $r$. The probability $B^m(r)$ shows the total number of $m$-dimensional matched vectors [2]. Similarly, $B^{m+1}(r)$ is defined for embedding dimension of $m+1$. Finally, the SampEn is defined as follows [2]:

$$\text{SampEn}(y, m, r) = -\ln\left(B^{m+1}(r)/B^m(r)\right) \quad (1)$$

#### 2.1.2 Fuzzy entropy (FuzEn)

In this case, for the time series $\mathbf{y} = \{y_1, y_2, ..., y_N\}$, embedding dimension $m$, and tolerance $r$, $U_t^m = \{y_t, y_{t+1}, ..., y_{t+m-1}\} - y0_t$ is formed where $y0_t = \sum_{j=0}^{m-1} \frac{y_{t+j}}{m}$. The distance between each of $U_{t_1}^m$ and $U_{t_2}^m$ is defined as $d_{t_1 t_2} = d\left[U_{t_1}^m, U_{t_2}^m\right] = \max\left\{\left|U_{t_1+k}^m - U_{t_2+k}^m\right| : 0 \leq k \leq m-1 \text{ and } t_1 \neq t_2\right\}$. Given FuzEn power $n$ and tolerance $r$, the similarity degree

$d_{t_1 t_2}$ is calculated through a fuzzy function $\mu(d_{t_1 t_2}, n, r)$ as $\exp(-(d_{t_1 t_2})^n / r)$. The function $\phi^m$ is then defined as

$$\phi^m(y, n, r) = \frac{1}{N-m} \sum_{t_1=1}^{N-m} \frac{1}{N-m-1} \sum_{t_2=1, t_1 \neq t_2}^{N-m} \exp(-(d_{t_1 t_2})^n / r) \quad (2)$$

Finally, the FuzEn of the signal is defined as the negative natural logarithm of the ratio of $\phi^m$ and $\phi^{m+1}$ (computed following the same procedure for embedding dimension $m + 1$) [3]:

$$\text{FuzEn}(y, m, n, r) = -\ln\left(\phi^{m+1} / \phi^m\right) \quad (3)$$

## 2.2 Coarse-graining for multiscale entropy

A "coarse-graining" process is applied to a time series $\{x_1, x_2, ..., x_b, ..., x_C\}$ where $C$ is the length of the signal. Each element of the coarse-grained time series for $\text{MSE}_\mu / \text{MFE}_\mu$ is defined as

$$^\mu y_i^{(\tau)} = \frac{1}{\tau} \sum_{b=(i-1)\tau+1}^{i\tau} x_b \quad 1 \leq i \leq \left\lfloor \frac{C}{\tau} \right\rfloor = N \quad (4)$$

where $\tau$ is the time scale factor [9]. This means that these coarse-grained sequences are computed as the average of consecutive samples. Costa et al. [12] also have recently proposed to use the variance, instead of the mean value, as follows:

$$^{\sigma 2} y_i^{(\tau)} = \frac{1}{\tau} \sum_{b=(i-1)\tau+1}^{i\tau} \left(x_b - {^\mu y_i^{(\tau)}}\right)^2, \quad 1 \leq i \leq \left\lfloor \frac{C}{\tau} \right\rfloor = N \quad (5)$$

The dimension of variance is not the same as the samples of the original signal, and the quadratic behavior of variance causes the differences between the data points and their corresponding average to become larger and smaller, respectively, for those differences which are larger and smaller than 1. To alleviate these shortcomings, we propose to use $\sigma$ in the coarse-graining process as a measure of spread via

$$^\sigma y_i^{(\tau)} = \sqrt{\frac{1}{\tau} \sum_{b=(i-1)\tau+1}^{i\tau} \left(x_b - {^\mu y_i^{(\tau)}}\right)^2}, \quad 1 \leq i \leq \left\lfloor \frac{C}{\tau} \right\rfloor = N \quad (6)$$

## 2.3 Refined composite multiscale fuzzy entropy

The traditional application of the coarse-graining procedure in $\text{MSE}_\mu$ leads to two main shortcomings. First, the $\text{MSE}_\mu$ is not symmetric in its dependency on the samples of the original time series. For example, in scale 3, we could rationally expect the measure to behave the same for $x_3$ and $x_4$, in comparison with $x_2$ and $x_3$. However, at scale 3, $x_1$, $x_2$, and $x_3$ are separated from $x_4$, $x_5$, and $x_6$. This phenomenon is illustrated in [15]. The second shortcoming is the variability of the entropy results for high-scale factors. When the $\text{MSE}_\mu$ is computed, the number of samples of the resulting coarse-grained sequence is $\lfloor C/\tau \rfloor = N$. When the scale factor $\tau$ is high, the number of time points in the coarse-grained sequence decreases. This may yield an unstable measure of entropy.

To alleviate these drawbacks, the improved multiscale permutation entropy and $\text{RCMSE}_\mu$ algorithm were proposed [13, 15]. Here, considering the advantages of FuzEn over SampEn, and $\text{RCMSE}_\mu$ over $\text{MSE}_\mu$, we introduce $\text{RCMFE}_\sigma$ and $\text{RCMFE}_\mu$.

The $\text{RCMFE}_\sigma$ is calculated in two main steps:

First, $z_u^{(\tau)} = \{y_{u,1}^{(\tau)}, y_{u,2}^{(\tau)}, ...\}$, $1 \leq u \leq \tau$ are generated, where $^\sigma y_{u,j}^{(\tau)} = \sqrt{\frac{1}{\tau} \sum_{b=u+\tau(j-1)}^{u+\tau j-1} \left(x_b - {^\mu y_{u,j}^{(\tau)}}\right)^2}$, where $^\mu y_{u,j}^{(\tau)} = \frac{\sum_{b=u+\tau(j-1)}^{u+\tau j-1} x_b}{\tau}$. In the $\text{RCMFE}_\sigma$ algorithm, for each scale factor $\tau$, we have $\tau$ different time series $z_u^{(\tau)}|(u = 1, ..., \tau)$, while in the MSE/MFE methods, only $z_1^{(\tau)}$ is considered [15].

For a defined scale factor $\tau$ and embedding dimension $m$, $\phi_{\tau,k}^m|(k = 1, ..., \tau)$ and $\phi_{\tau,k}^{m+1}|(k = 1, ..., \tau)$ for each of $z_k^{(\tau)}|(k = 1, ..., \tau)$ are separately calculated. Next, the average of values of $\phi_{\tau,k}^m$ and $\phi_{\tau,k}^{m+1}$ on $1 \leq k \leq \tau$ are computed, respectively. Finally, the $\text{RCMFE}_\sigma$ is computed as follows:

$$\text{RCMFE}_\sigma(x, \tau, m, n, r) = -\ln\left(\overline{\phi_\tau^{m+1}} / \overline{\phi_\tau^m}\right) \quad (7)$$

It should be mentioned that the difference between $\text{RCMFE}_\sigma$ and $\text{RCMFE}_\mu$ is that the latter one uses $^\mu y_{u,j}^{(\tau)} = \frac{\sum_{b=u+\tau(j-1)}^{u+\tau j-1} x_b}{\tau}$, whereas the first one uses Eq. 6 in their first step of algorithm. The embedding dimension $m$, FuzEn power $n$, and tolerance $r$ for all of the approaches were respectively chosen as 2, 2, and 0.15 multiplied by the standard deviation of the original time series [2, 3, 16].

## 2.4 Evaluation signals

### 2.4.1 Noise and synthetic signals

In this subsection, the signals used to study the mentioned multiscale approaches and their interpretability in terms of classical signal processing concepts are described.

First, we consider the performance of the multiscale entropy metrics on WGN and $1/f$ noise. The number of sample points of each of the WGN and $1/f$ noise was 40,000. In addition, we consider other synthetic signals with a sampling frequency ($f_s$) of 150 Hz and a length of 100 s (15,000 sample points). The time plots of these synthetic signals, and their corresponding spectrograms, and two zooms (for each kind of signal) on their start and end, to show the changes in their characteristics, are illustrated in Fig. 1. All of them have been employed to inspect the Lempel-Ziv complexity measure, improved permutation entropy, or auto-mutual information function rate of decrease and have been described in [15, 17, 18], respectively, where additional details can be found.

1. RCMFE$_\sigma$ and RCMFE$_\mu$ versus noise: The dependency between the abovementioned multiscale entropy-based methods and $1/f$ noise and WGN is considered in this paper. WGN has a constant power spectral density as WGN is a signal whose samples are randomly drawn from a Gaussian distribution and uncorrelated [19]. The power spectral density of a stochastic process appropriate to model evolutionary or developmental systems is characterized by equal energy per octave as $1/f$ noise [20].

2. RCMFE$_\sigma$ and RCMFE$_\mu$ versus frequency: In order to clarify how the RCMFE$_\sigma$/RCMFE$_\mu$ changes when the frequency of sinusoidal signals varies, a constant amplitude chirp signal whose frequency is swept logarithmically from 0.1 to 30 Hz in 100 s is considered [15, 17]. RCMFE$_\sigma$ and the other multiscale entropy methods were applied to this signal using a moving window of 2000 samples (13.33 s) with 90% overlap. Fig. 1a demonstrates the constant chirp signal.

3. RCMFE$_\sigma$ and RCMFE$_\mu$ versus spectral content of colored noise: In order to find the relationship between the RCMFE$_\sigma$ or RCMFE$_\mu$ and the spectral content of colored noise, an autoregressive (AR) process of order 1, $AR(1)$, was generated varying the model parameter, $\rho$, linearly from +0.9 to −0.9. Its energy hence moved from low to high frequencies. In case of $\rho = 0$, the sequence corresponded to WGN, in the center of the synthetic time series. Fig. 1b shows the corresponding spectrogram, time plot, and zoom views.

4. RCMFE$_\sigma$ and RCMFE$_\mu$ versus changes from randomness to orderliness: In order to consider how the RCMFE$_\sigma$ and RCMFE$_\mu$ change when a stochastic sequence progressively turns into a periodic deterministic time series, we created a MIX process employed in [18, 21, 22]. This is defined as follows:

$$\mathrm{MIX} = (1-z)x + zy \tag{8}$$

where $z$ denotes a random variable which is equal to 1 with probability $p$ and is equal to 0 with probability $1 - p$, $x$ depicts a periodic synthetic signal as $x_k = \sqrt{2}\sin(2\pi k/12)$, and $y$ is a uniformly distributed variable on $[-\sqrt{3}, \sqrt{3}]$ [18, 21]. Thus, the lower $p$ is selected, the more regular or periodic the time series is, while higher $p$ leads to more irregular signal. In this sense, to show the evolution from randomness to orderliness, $p$ is linearly changed from 0.01 to 0.99. This signal is depicted in Fig. 1c.

5. RCMFE$_\sigma$ and RCMFE$_\mu$ versus changes from periodicity to non-periodic non-linearity: In order to clarify the dependence of the multiscale entropies on these changes, the logistic map is employed. This analysis is dependent on the model parameter $\alpha$ [18, 21] as follows:

$$x_k = \alpha x_{k-1}(1-x_{k-1}) \tag{9}$$

The synthetic signal $x$ was created varying the parameter $\alpha$ linearly from 3.5 to 3.99. With $\alpha = 3.5$, the signal oscillated among four values. For $\alpha$ between 3.5 and 3.57, the signal is periodic and the number of values doubles progressively. For $3.57 \leq \alpha \leq 3.99$, the time series is chaotic, although it has windows of periodic behavior (e.g., $\alpha \approx 3.8$, as seen in Fig. 1d) [23].

6. RCMFE$_\sigma$ and RCMFE$_\mu$ versus different non-linear regimes: In order to investigate the changes in the behavior of a non-linear system, the Lorenz attractor is used here as

$$\begin{aligned}
\dot{x} &= \lambda(y-x) \\
\dot{y} &= x(\rho-z)-y \\
\dot{z} &= xy-\beta z
\end{aligned} \tag{10}$$

where $\lambda$, $\beta$, and $\rho$ denote the system parameters [23, 24]. The first segment of this time series has a length of 7500 sample points, and it was created with $\lambda = 10$, $\beta = 8/3$, and $\rho = 28$. Therefore, it has a chaotic behavior. The second segment, which has 7500 sample points, was generated with $\lambda = 10$, $\beta = 8/3$, and $\rho = 99.96$. It exhibits a torus knot [17, 23]. Both segments were created by the use of a fixed step-size first-order integration technique without pre-integration and with the step size set to $1/f_s$. It should be noted that these two segments were normalized with standard deviation (SD) of 1, after these segments had been generated. The coordinate $x$, which is the signal analyzed in this article, appears in Fig. 1e.

### 2.4.2 Clinical datasets

The ability of the newly proposed RCMFE$_\mu$ and RCMFE$_\sigma$ to distinguish different types of physiological activity was tested on the following clinical datasets: MEG resting state activity in AD and EEG signals of focal and non-focal origin in epilepsy.

The MEG signals were acquired utilizing a 148-channel whole-head magnetometer (Magnes 2500 WH, 4D Neuroimaging) located in a magnetically shielded room at the

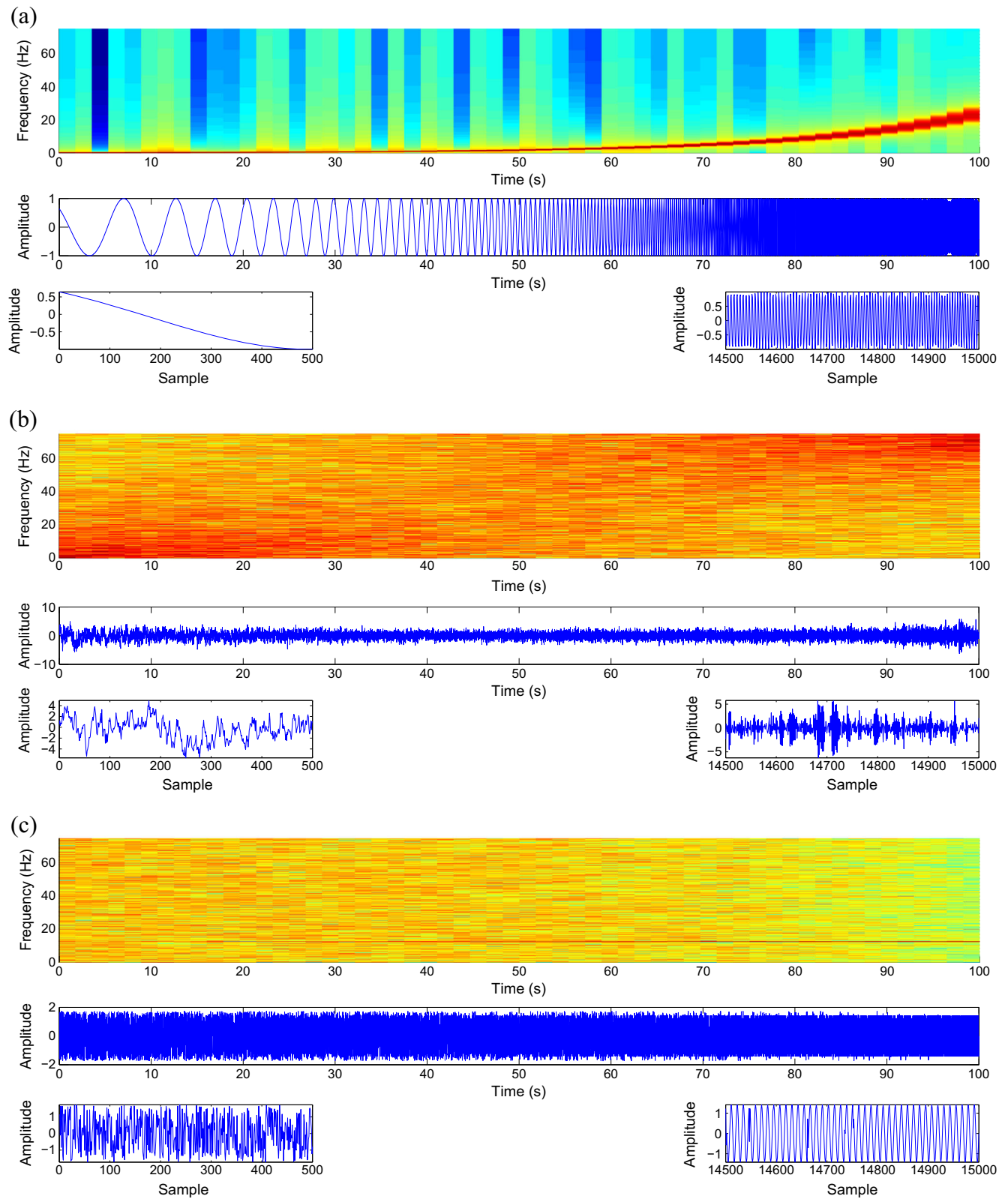

**Fig. 1** Spectrograms, time plots, and zoom views on the first and last time intervals of the synthetic signals used in this study. **a** Chirp signal with constant amplitude. **b** AR(1) process with variable parameter $\rho$. **c** MIX process evolving from randomness to periodic oscillations. **d** Logistic map signal. **e** Lorenz system with two different non-linear dynamics. *Red* corresponds to high power and *blue* corresponds to low power (color figure online)

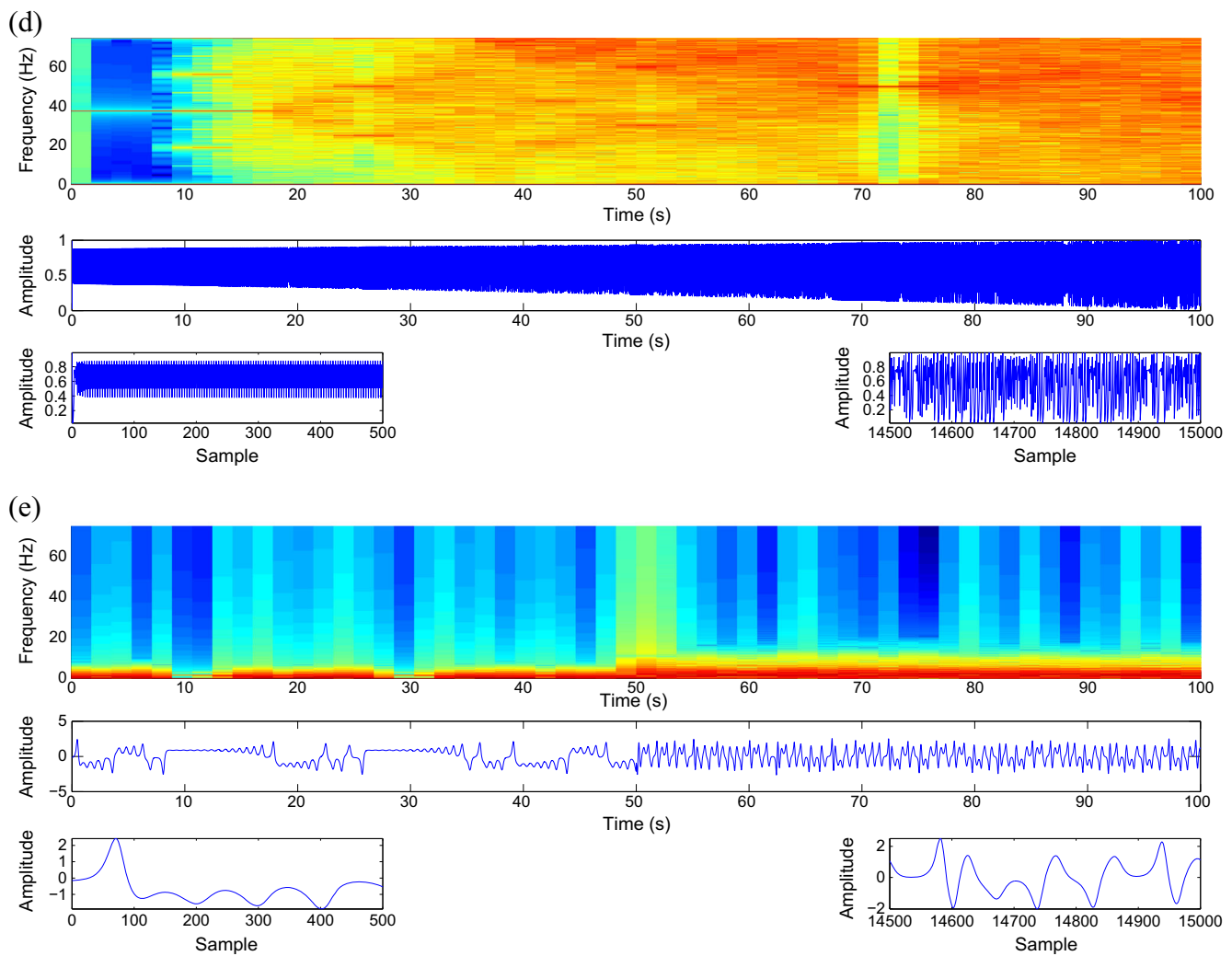

**Fig. 1** (continued)

"Centro de Magnetoencefalografia Dr. Perez-Modrego," Spain. Resting-state MEG activity was recorded from 36 patients with probable AD [25] (24 women; age = 74.06 ± 6.95 years, mean ± standard deviation; MMSE score = 18.06 ± 3.36) and 26 age-matched controls (17 women; age = 71.77 ± 6.38 years; MMSE score = 28.88 ± 1.18). The subjects laid on a hospital bed in a relaxed state with eyes closed. For each participant, 5 min of MEG resting-state activity was recorded at a sampling frequency ($f_s$) of 169.54 Hz. The signals were divided into segments of 10s (1695 samples per channel) and visually inspected using an automated thresholding procedure to discard segments significantly contaminated with artifacts [26]. The effect of cardiac artifact was reduced from the recordings using a constraint blind source separation procedure. Finally, a band-pass FIR filter with cutoffs at 1.5 and 40 Hz was applied to the data. For more information about the dataset, please refer to [27]. For each subject and each channel, we analyzed each epoch of 10s individually and the average of results is reported. Note that all control subjects and AD patients' caregivers gave informed consent for participation in the study, which was approved by the local Ethics Committee [27].

The intracranial EEG signals were recorded from five patients suffering from pharmacoresistant focal-onset epilepsy leading to two main separate sets of signals. The first one was recorded from brain regions where the primarily ictal EEG recording changes were detected as judged by expert visual inspection ("focal signals"). The second set of signals was recorded from brain regions not involved at seizure onset ("non-focal signals"). Each set includes five patients. Each patient consists of 750 pair signals, and the length of each of them was 10,240 sample points or 20 s. The sampling frequency was 512 Hz. Each pair includes two EEG time series which are recorded from adjacent channels which here we consider the first time series. They also provided a subset of the recordings containing the first 50 signals for each set. We use this subset to evaluate the proposed methods. For more information about the dataset, please refer to [28]. Before computing the multiscale entropy approaches, all signals were

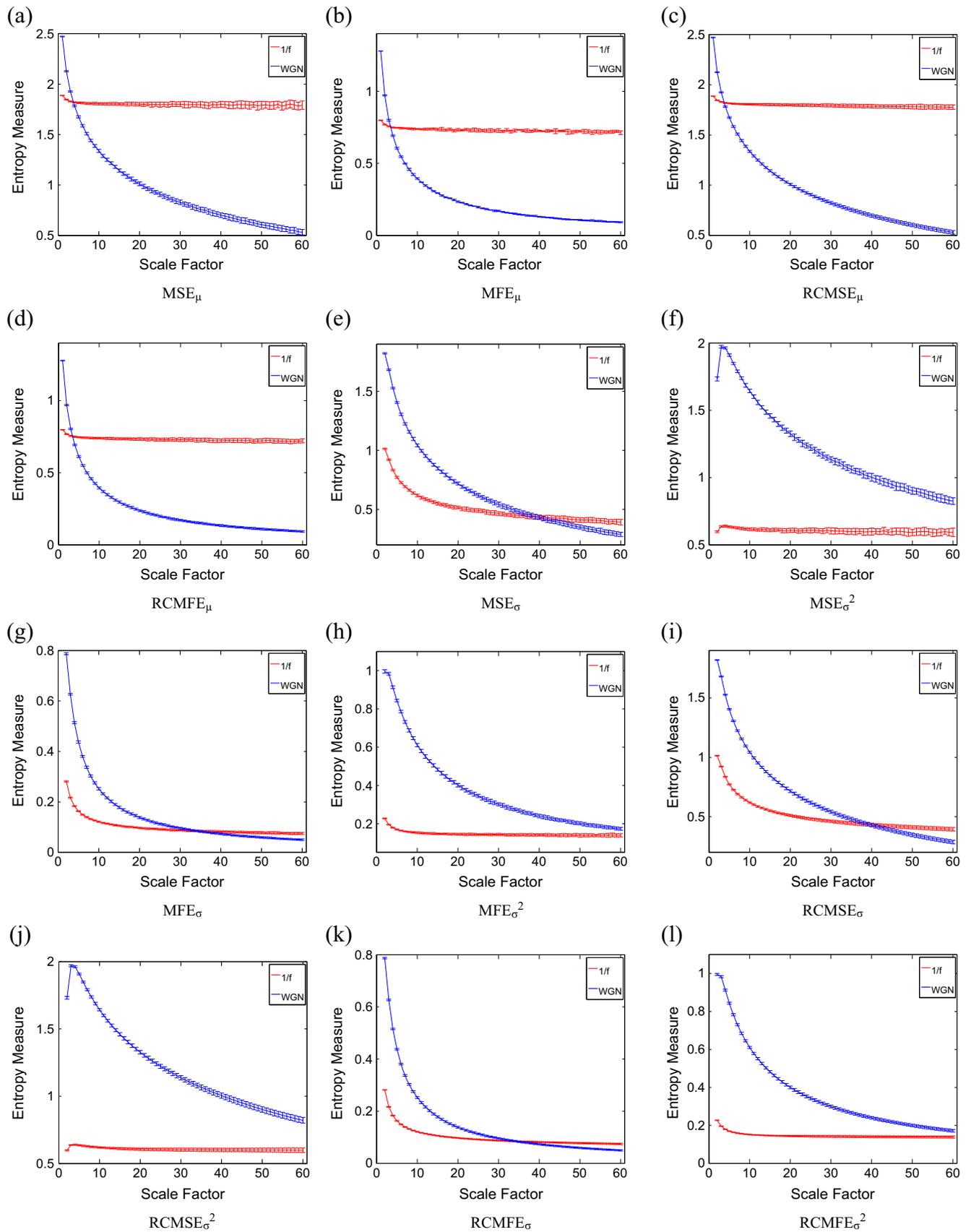

◀ **Fig. 2** Mean value and SD of results of the **a** $MSE_\mu$, **b** $MFE_\mu$, **c** $RCMSE_\mu$, **d** $RCMFE_\mu$, **e** $MSE_\sigma$, **f** $MSE_\sigma^2$, **g** $MFE_\sigma$, **h** $MFE_\sigma^2$, **i** $RCMSE_\sigma$, **j** $RCMSE_\sigma^2$, **k** $RCMFE_\sigma$, and **l** $RCMFE_\sigma^2$ computed from 40 different $1/f$ noise test signals. Red and blue indicate $1/f$ noise and WGN results, respectively digitally filtered employing an FIR band-pass filter with cutoff frequencies at 0.5 and 40 Hz. Note that retrospective EEG data analysis has been approved by the ethics committee of the Kanton of Bern. Moreover, all patients gave written informed consent that the obtained signals from long-term EEG might be utilized for research purposes [28].

## 3 Results

### 3.1 Noise signals

First, we consider WGN and $1/f$ noise as two widely used signals tested in multiscale entropy methods [8, 13]. The results for $MSE_\mu$, $MFE_\mu$, $RCMSE_\mu$, $RCMFE_\mu$, $MSE_\sigma$, $MSE_\sigma^2$, $MFE_\sigma$, $MFE_\sigma^2$, $RCMSE_\sigma$, $RCMSE_\sigma^2$, $RCMFE_\sigma$, and $RCMFE_\sigma^2$ are depicted in Fig. 2a–l, respectively. As it can be observed in Fig. 2, for WGN, the entropy values of all multiscale approaches, except $MSE_\sigma^2$ and $RCMSE_\sigma^2$, decrease monotonically with scale factor $\tau$. However, for $1/f$ noise, the entropy values become approximately constant over larger-scale factors. These facts are in agreement with WGN which only has structure in the shortest temporal scale, whereas $1/f$ noise has structure across all scales [8, 13]. Note that each error bar of each scale factor $\tau$ depicts the SD of the results of 40 signals for each WGN or $1/f$ noise.

Comparing results obtained by $MSE_\mu$ (Fig. 2a) and $MFE_\mu$ (Fig. 2b) shows, as expected theoretically, that the $MFE_\mu$ leads to a smaller variability in the results. Statistical tests confirmed the smaller variability of the $MFE_\mu$ results ($p$ value $\leq 0.05$) as assessed with Levene's test at $\tau = 60$. In addition, the $RCMSE_\mu$/$RCMFE_\mu$ profiles have smaller SDs than $MSE_\mu$/$MFE_\mu$.

Although the $MSE_\sigma^2$ values for WGN are larger than $1/f$ noise for scale factors 1 to 60, according to Fig. 2f, it is predicted that this measure for WGN will become smaller than those of $1/f$ noise for large enough scale factors. For $MSE_\sigma$ and for scale factors 1 to 37, the larger entropy values are assigned to WGN signal in comparison with $1/f$ noise, while for scale factors larger than 37, the SampEn values for $1/f$ noise are larger than those of WGN, in agreement with the fact that $1/f$ noise is considered more structurally complex across multiple scales [9, 29]. Comparing the results shows that crossing between WGN and $1/f$ noise does not happen at short levels of scale factor for the coarse-graining process based on variance and standard deviation, unlike the mean.

It should be added that the results obtained for parameter $r$, used in [12], are similar to our results with $r = 0.15$ multiplied by the SD of that time series, employed in [16].

In order to understand the importance of refined composite technique on the basic multiscale entropy methods, we employed the coefficient of variation (CV) defined as the SD divided by the mean [30]. The main purpose to employ such a measure is that the SDs of data may increase or decrease proportionally to the mean. Thus, the CV, as a standardization of the SD, permits comparison of variability estimates regardless of the magnitude of the variable [30]. We study the results for $1/f$ noise and WGN signals at scale factor 20. As can be seen in Table 1, the refined composite technique decreases the CV values of the basic multiscale approaches, leading to more stable results.

The computation times of the conventional and proposed multiscale sample and fuzzy entropy approaches with the maximum scale factor 60 for the WGN signals with the length of 40,000 sample points are demonstrated in Table 2. The simulations have been carried out using a PC with Intel® Xeon® CPU, E5420, 2.5 GHz, and 8-GB RAM by MATLAB R2010a. The results show that FuzEn-based methods are slower than SampEn-based ones and the refined composite technique increases the computation time significantly. The running times of the variance-based methods are similar to those of the standard deviation-based algorithms. Moreover, since the $MSE_\sigma^2$, $MSE_\sigma$, $MFE_\sigma^2$, $MFE_\sigma$, $RCMSE_\sigma^2$, $RCMSE_\sigma$, $RCMFE_\sigma^2$, and $RCMFE_\sigma$ start from scale factor 2 and the computation cost of SampEn and FuzEn is $O(N^2)$ [31], the running times of these kinds of algorithms are noticeably smaller than those of the algorithms based on coarse-graining with regard to the mean.

### 3.2 Sensitivity of multiscale methods to signal length

To evaluate the sensitivity of multiscale methods to the signal length, we consider WGN and $1/f$ noise signals as functions of sample points size $C$. Figures 3, 4, 5, and 6 respectively depict the $MSE_\mu$, $RCMSE_\mu$, $MFE_\mu$, and $RCMFE_\mu$ values for the signal length 100, 300, 1000, 3000, 10,000, and 30,000 computed from 40 different realizations of WGN and $1/f$ noise. The results show that the greater the value of $C$, the more robust the multiscale entropy estimations, as seen from the error bars.

**Table 1** The CV values of the proposed and classical multiscale entropy-based analyses at scale factor 20 for $1/f$ noise and WGN

| Signals | $MSE_\mu$ | $MFE_\mu$ | $RCMSE_\mu$ | $RCMFE_\mu$ |
|---|---|---|---|---|
| $1/f$ | 0.015 | 0.013 | 0.011 | 0.011 |
| WGN | 0.019 | 0.019 | 0.011 | 0.010 |
|  | $MSE_\sigma$ | $MFE_\sigma$ | $RCMSE_\sigma$ | $RCMFE_\sigma$ |
| $1/f$ | 0.023 | 0.023 | 0.017 | 0.016 |
| WGN | 0.022 | 0.020 | 0.020 | 0.015 |
|  | $MSE_\sigma^2$ | $MFE_\sigma^2$ | $RCMSE_\sigma^2$ | $RCMFE_\sigma^2$ |
| $1/f$ | 0.026 | 0.025 | 0.017 | 0.016 |
| WGN | 0.015 | 0.018 | 0.010 | 0.010 |

**Table 2** Computation time of the classical and proposed multiscale sample and fuzzy entropy methods

| $MSE_\mu$ | $MFE_\mu$ | $RCMSE_\mu$ | $RCMFE_\mu$ |
|---|---|---|---|
| 49.08 s | 73.21 s | 253.61 s | 364.73 s |
| $MSE_{\sigma}^2$ | $MFE_{\sigma}^2$ | $RCMSE_{\sigma}^2$ | $RCMFE_{\sigma}^2$ |
| 23.08 | 35.79 s | 186.99 s | 299.03 s |
| $MSE_\sigma$ | $MFE_\sigma$ | $RCMSE_\sigma$ | $RCMFE_\sigma$ |
| 22.94 s | 34.92 s | 189.24 s | 282.62 s |

It has been suggested that the number of sample points is at least $10^m$, or preferably at least $30^m$, to robustly estimate approximate entropy or SampEn in time series [32]. Because the coarse-graining step reduces the times series length by the scale factor $\tau$, and here we have $\tau_{\max} = 10$ and $m = 2$, the original signal should have at least 1000 samples. As mentioned before, in SampEn, the number of instances where $d\left[Y_{t_1}^m, Y_{t_2}^m\right]$ is smaller than a predefined tolerance $r$ is counted. If the length of a time series is too small, this number may be 0, leading to an undefined entropy measure. According to this fact, the results obtained by $MSE_\mu$ for $C = 100$ and 300, respectively depicted in Fig. 3a, b, are undefined.

For $RCMSE_\mu$ at scale factor $\tau$, although the length of the signal decreases $\tau$ times, we take into account $\tau$ time coarse-grained signals, instead of only one signal as in conventional multiscale entropy approaches [13]. Therefore, in refined composite-based algorithms, we have $\tau$ times more number of instances in comparison with their corresponding basic versions, leading to more reliable results, especially for short signals. This fact can be seen in Fig. 4 in comparison with Fig. 3. Although $RCMSE_\mu$ outperforms $MSE_\mu$ in terms of reliability for short signals, $RCMSE_\mu$ values for $C = 100$ and $C = 300$ (Fig. 4a, b) are still undefined at some scale factors.

However, the FuzEn-based algorithms do not count matches, yet consider all possible range of distances between any two composite vectors. Therefore, $MFE_\mu$ and $RCMFE_\mu$ avoid resulting in undefined entropy values in such situations. The results obtained by the $RCMFE_\mu$ (Fig. 6) have considerably smaller SD values, especially for short signals, than those obtained by $MFE_\mu$ (Fig. 5).

### 3.3 Synthetic signals

To understand the effect of frequency on multiscale entropy-based methods, we employed a sliding window moving along each of the abovementioned synthetic signals. Then, for each scale factor, the multiscale entropy-based method of that part of the signal was computed. Because the length of the window is 2000 sample points, we consider the scale factor from 1 to 15, to ensure the length of the coarse-grained signals is enough for $m = 2$ [33].

For chirp signal with constant amplitude, the $RCMFE_\sigma$, $RCMFE_\mu$, $MSE_{\sigma}^2$, and $MSE_\mu$ results are respectively

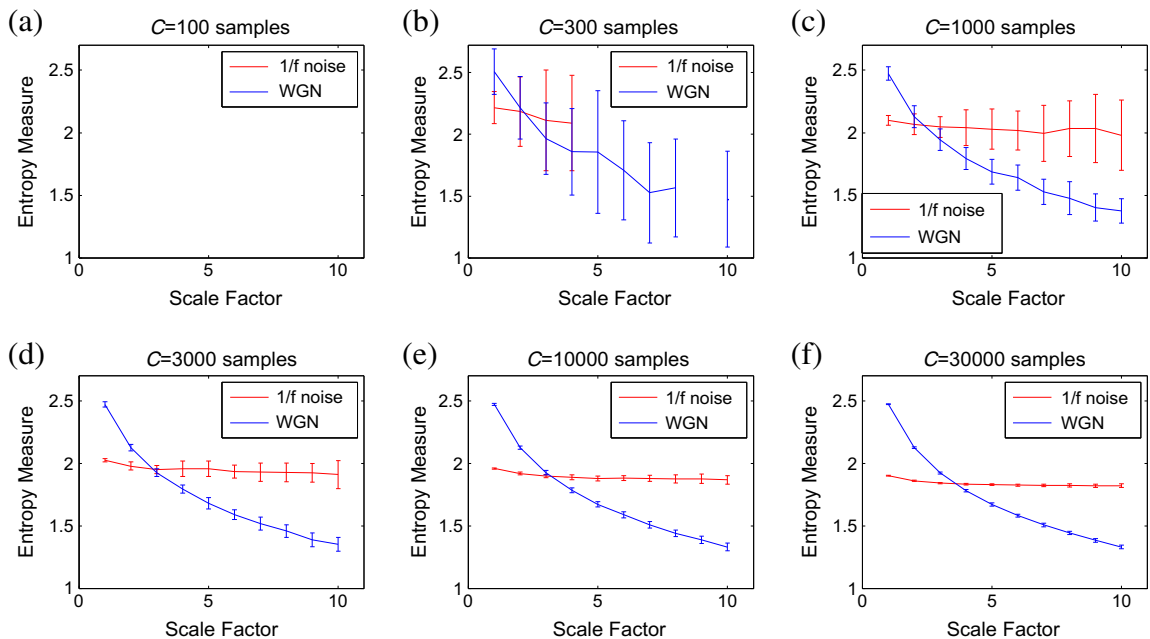

**Fig. 3** $MSE_\mu$ as a function of data length $C$, **a** $C = 100$, **b** $C = 300$, **c** $C = 1000$, **d** $C = 3000$, **e** $C = 10,000$, and **f** $C = 30,000$ computed from 40 different WGN and $1/f$ noise signals. The entropy values are undefined for noise signals with the length of 100 and 300 at all and large-scale factors, respectively. *Red* and *blue* demonstrate $1/f$ noise and WGN results, respectively (color figure online)

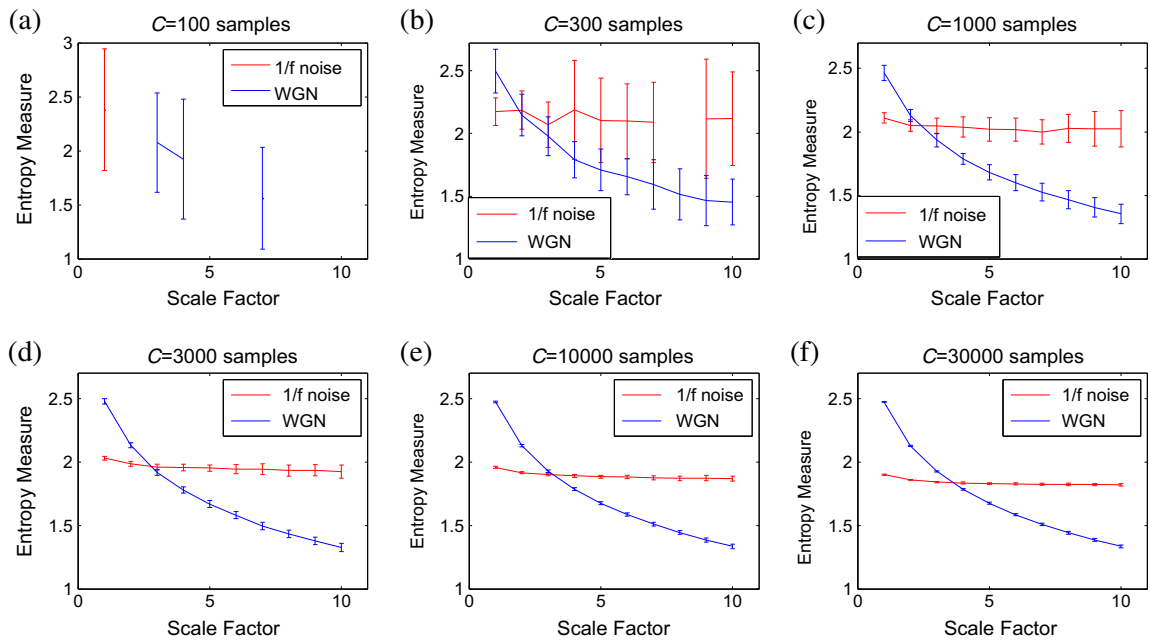

**Fig. 4** $RCMSE_\mu$ as a function of data length $C$, **a** $C = 100$, **b** $C = 300$, **c** $C = 1000$, **d** $C = 3000$, **e** $C = 10{,}000$, and **f** $C = 30{,}000$ computed from 40 different WGN and $1/f$ noise signals. The entropy values are undefined for noise signals with the length of 100 and 300 at all and large-scale factors, respectively. *Red* and *blue* demonstrate $1/f$ noise and WGN results, respectively (color figure online)

shown in Fig. 7a–d. When the time window is occupied at the beginning of the signal, which has smaller frequency, the FuzEn and SampEn values are low across all $\tau$. As expected theoretically, all the $RCMFE_\sigma$, $RCMFE_\mu$, $MSE_\sigma^2$, and $MSE_\mu$ values increase with higher frequencies, which happens in later temporal windows (TWs). It is worth noting that since the SD/variance, unlike the mean value, of one single number is 0, the entropy measure in the first scale factor is undefined. This fact can be seen in Fig. 7a, c in comparison with Fig. 7b, d.

In Fig. 7e–h, it can be observed generally, using an $AR(1)$ process with variable parameter, that the entropy measures of $RCMFE_\sigma$, $MFE_\sigma^2$, and $MFE_\sigma$, unlike $RCMFE_\mu$, increase in higher TWs in every scale factor.

Figure 7i–l respectively shows the results obtained by $RCMFE_\sigma$, $RCMFE_\mu$, $MFE_\sigma$, and $MFE_\mu$ using the

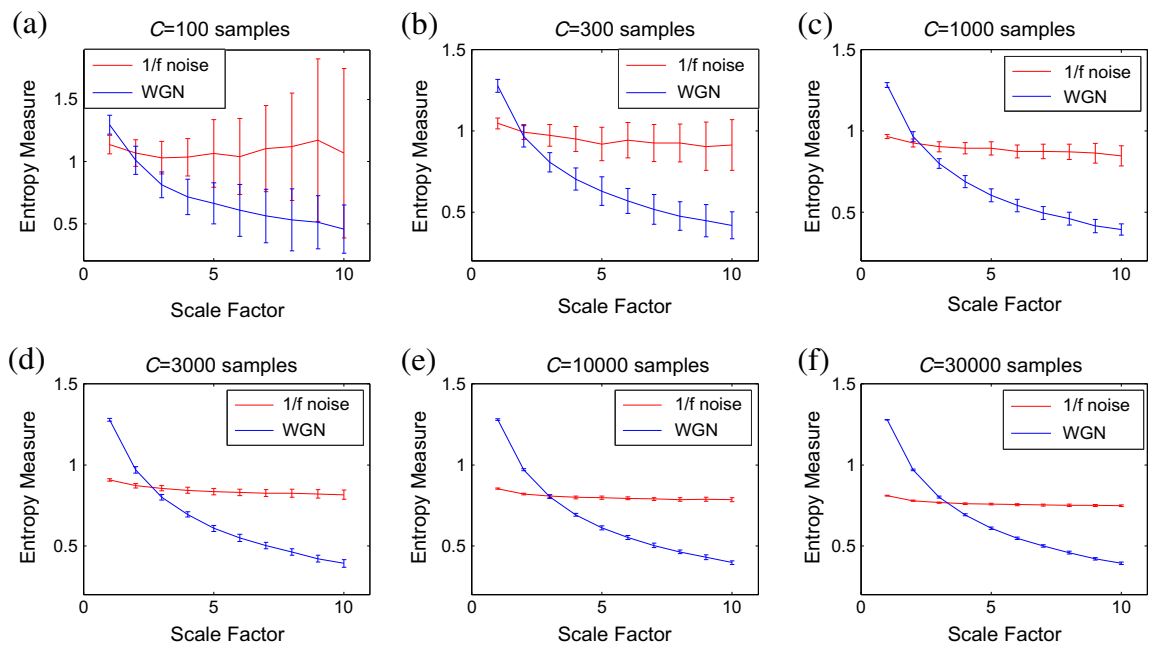

**Fig. 5** $MFE_\mu$ as a function of data length $C$, **a** $C = 100$, **b** $C = 300$, **c** $C = 1000$, **d** $C = 3000$, **e** $C = 10{,}000$, and **f** $C = 30{,}000$ computed from 40 different WGN and $1/f$ noise signals. *Red* and *blue* demonstrate $1/f$ noise and WGN results, respectively (color figure online)

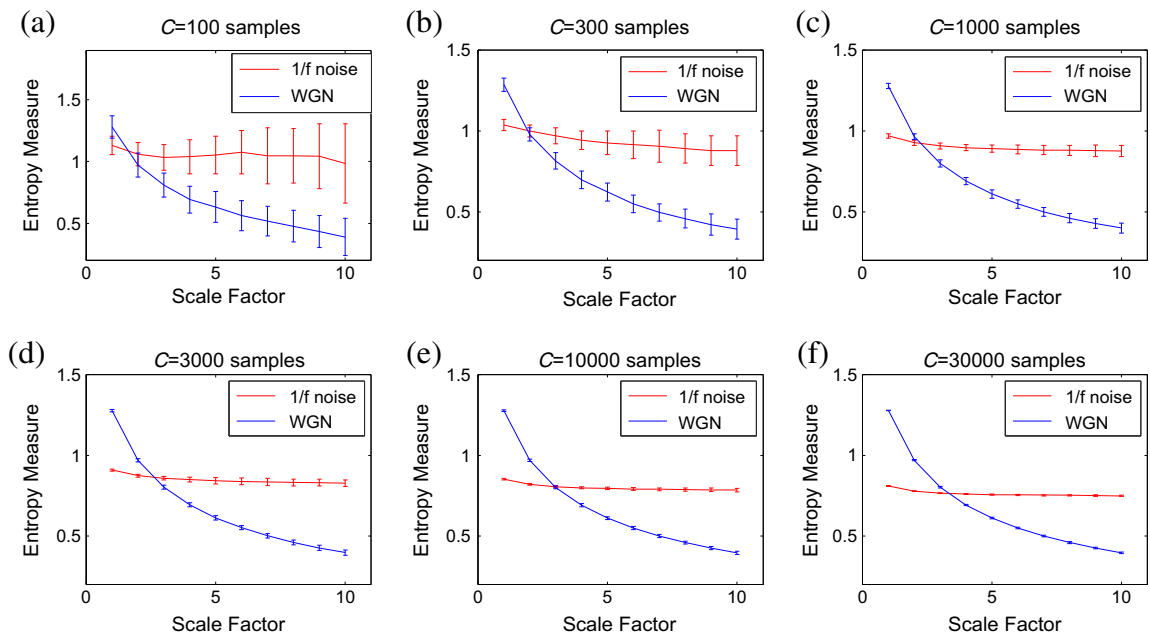

**Fig. 6** $RCMFE_\mu$ as a function of data length $C$, **a** $C = 100$, **b** $C = 300$, **c** $C = 1000$, **d** $C = 3000$, **e** $C = 10{,}000$, and **f** $C = 30{,}000$ computed from 40 different WGN and $1/f$ noise signals. *Red* and *blue* demonstrate $1/f$ noise and WGN results, respectively (color figure online)

abovementioned MIX process. The entropy measures of all of them decrease in higher TWs in every scale factor, showing the evolution from randomness to periodic oscillations.

Figure 7m–p illustrates the results obtained by $RCMFE_\sigma$, $RCMFE_\mu$, $MSE_\sigma$, and $MSE_\mu$, respectively, using the logistic map which the parameter $\alpha$ changes linearly from 3.5 to 3.99. The entropy measures, obtained by all of them, generally increase along the signal, at each scale factor, except for the downward spikes in the windows of periodic behavior. This fact is in agreement with Fig. 4.10 (page 87 in [23]). It is also supported by Fig. 1d which shows that the frequency of the signal for $t = 70$–$75$ s is lower than for its adjacent time samples. In case of increasing scale factor, the $RCMFE_\sigma$ and $MSE_\sigma$ results decrease, whereas the $RCMFE_\mu$ and $MSE_\mu$ results first increase respectively until $\tau = 2$ and $\tau = 4$ then decrease. It shows that mean- and standard deviation-based multiscale approaches, extracting different kinds of dynamical properties of, respectively, mean and spread over multiple time scales, lead to different kinds of features.

Using the Lorentz system, we find that $RCMFE_\sigma$, $RCMFE_\mu$, $MSE_\mu$, and $RCMSE_\mu$ respectively shown in Fig. 7q–t can distinguish two different non-linear dynamics.

### 3.4 Clinical datasets

We also assess the suitability of the $RCMFE_\mu$ and $RCMFE_\sigma$ methods to characterize AD in MEG signals. The profiles are shown in Fig. 8. The average of $RCMFE_\sigma$ values for AD patients is smaller than that for controls at all scale factors. This is in agreement with [5, 34]. In contrast, the average of $RCMFE_\mu$ values for AD patients is smaller than that for controls for only $1 \leq \tau \leq 3$.

False discovery rate (FDR)-adjusted [35] $p$ values of a Student $t$ test assuming unequal variances for each MEG channel and temporal scale factor to evaluate the differences between the values of entropy for AD patients and controls are shown in Fig. 8 in a logarithmic scale. The FDR-adjusted $p$-values obtained by $RCMFE_\mu$, unlike those of $RCMFE_\sigma$, initially increase and then decrease along the temporal scale factor for almost all channels.

We also classify the AD subjects and controls using a naive Bayes classifier [36]. For each individual, 15 and 14 features (temporal scale factors) are extracted by averaging the $RCMFE_\mu$ and $RCMFE_\sigma$ results across all channels, respectively. We ran 200 repetitions of a tenfold cross-validation. The average classification accuracies were 72.81 and 78.22%, respectively, for $RCMFE_\mu$ and $RCMFE_\sigma$. This shows that, in this case, $RCMFE_\sigma$ features lead to higher classification accuracy than $RCMFE_\mu$ ones. The classification was done with the WEKA data mining software [37].

We also study the behavior of $RCMFE_\mu$ and $RCMFE_\sigma$ in focal and non-focal EEG time series. The error bars illustrating the distributions of the $RCMFE_\mu$ and $RCMFE_\sigma$ values computed from focal and non-focal EEG signals are shown in Fig. 9a, b until scale factor 30. For each scale factor, the average of entropy values of focal EEG signals is smaller than that of non-focal ones. It illustrates that the non-focal EEG recordings are generally more complex than the focal ones, and it is in agreement with [28] and [38].

We adjusted the FDR independently for each of $RCMFE_\mu$ and $RCMFE_\sigma$. The adjusted $p$ values are depicted in Fig. 9c, d

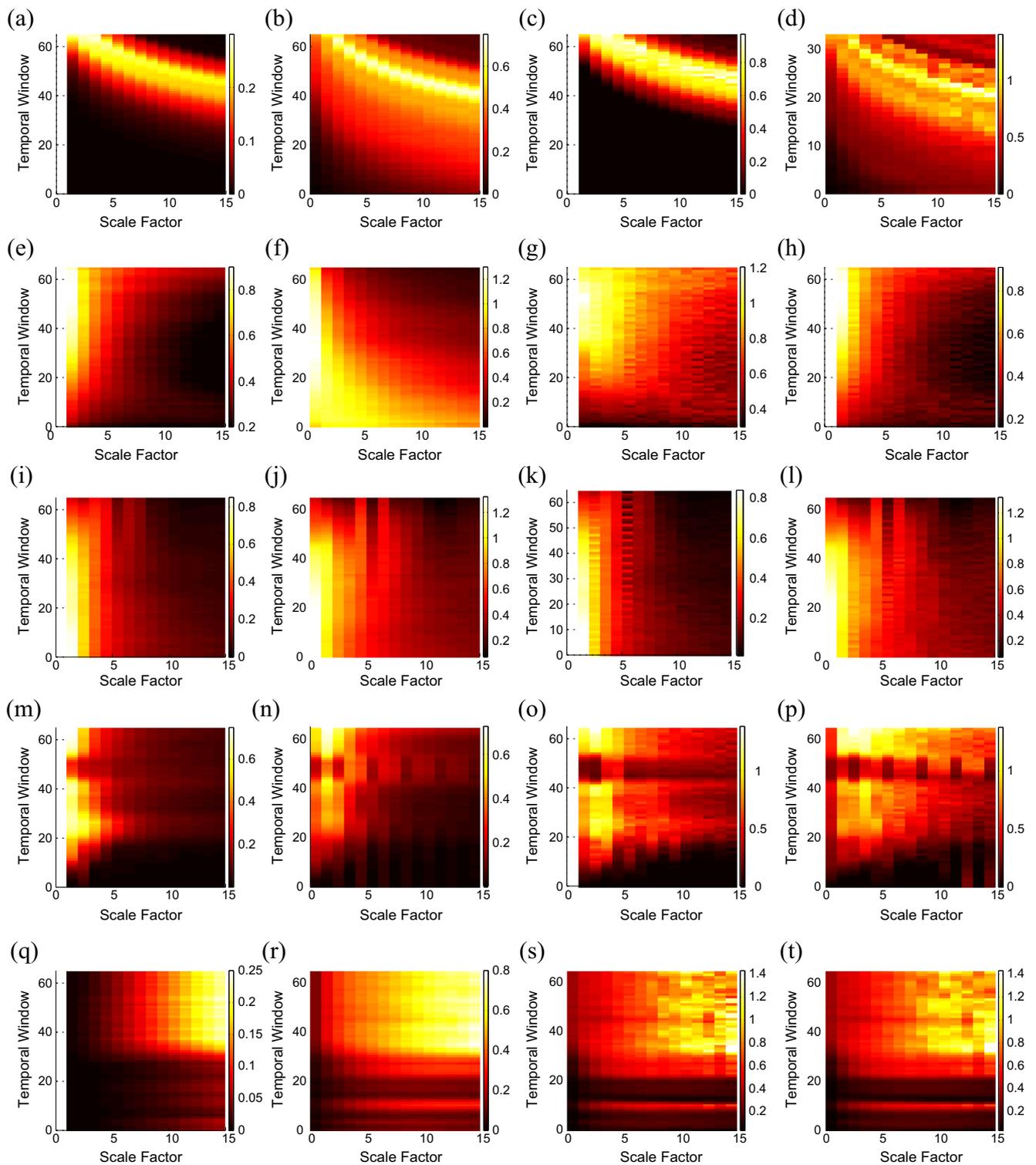

**Fig. 7** Results of the tests performed to understand better diverse multiscale entropy approaches and their interpretation. Relationships between chirp signal with constant amplitude and **a** RCMFE$_\sigma$, **b** RCMFE$_\mu$, **c** MSE$_\sigma^2$, and **d** MSE$_\mu$. Relationships between $AR(1)$ process with variable parameter and **e** RCMFE$_\sigma$, **f** RCMFE$_\mu$, **g** MFE$_\sigma^2$, and **h** MFE$_\sigma$. Relationships between the abovementioned MIX process and **i** RCMFE$_\sigma$, **j** RCMFE$_\mu$, **k** MFE$_\sigma$, and **l** MFE$_\mu$. Relationships between the logistic map and **m** RCMFE$_\sigma$, **n** RCMFE$_\mu$, **o** MSE$_\sigma$, and **p** MSE$_\mu$. Relationships between Lorenz system with two different non-linear dynamics and **q** RCMFE$_\sigma$, **r** RCMFE$_\mu$, **s** MSE$_\mu$, and **t** RCMSE$_\mu$

for RCMFE$_\mu$ and RCMFE$_\sigma$, respectively. The results show that the RCMFE$_\mu$ method achieves smaller adjusted $p$ values at scale factors 1–9, whereas the RCMFE$_\sigma$ algorithm leads to smaller adjusted $p$ values at scale factors 10–30,

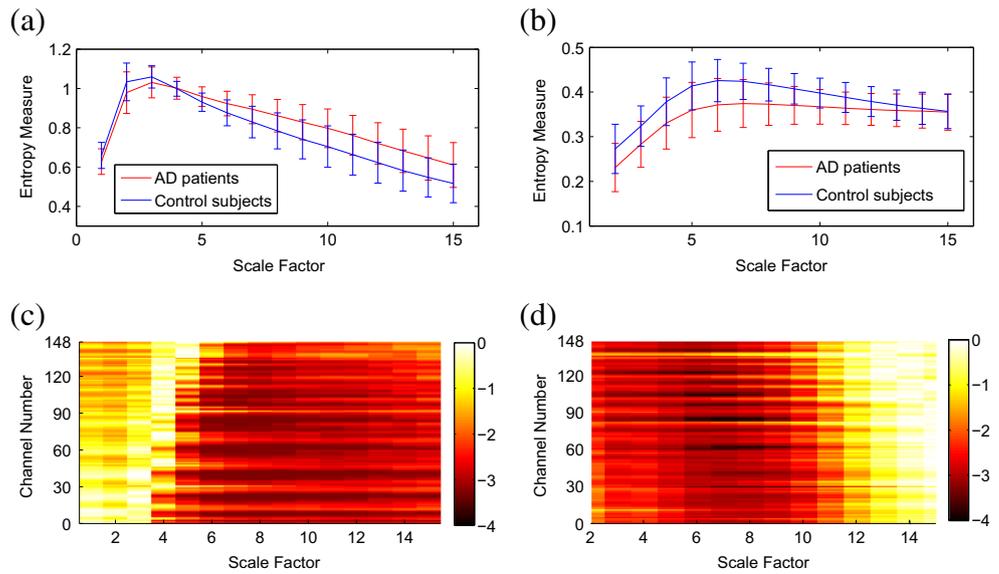

**Fig. 8** Plots illustrating the mean ± SD (as error bars) of the **a** RCMFE$_\mu$ and **b** RCMFE$_\sigma$ values for AD subjects and control subjects. Base-10 logarithm of the FDR-adjusted $p$ values for the differences in **c** RCMFE$_\mu$ and **d** RCMFE$_\sigma$ at each channel and temporal scale between AD patients and controls

demonstrating that when (RC)MFE$_\mu$ at specific scale factors cannot distinguish different kinds of dynamics, the (RC)MFE$_\sigma$ may do so and vice versa.

We also applied the same classification scheme to distinguish the focal and non-focal signals. The average classification accuracies were 71.58 and 79.62%, respectively, for RCMFE$_\mu$ and RCMFE$_\sigma$. It again shows that the RCMFE$_\sigma$ may lead to different or sometimes more useful information for characterization of signals.

To compare the existing and proposed univariate multiscale methods, we use FDR-adjusted $p$ values for focal versus non-focal signals as well as AD patients' versus controls' recordings. The results for scale factor 10 are shown in Table 3. The results demonstrate that standard deviation-based methods discriminate two groups for both the datasets better than variance- and mean-based multiscale algorithms. The adjusted $p$ values show that for clinical filtered data, unlike noisy time series, the refined composite technique does not improve the performance of the basic multiscale approaches noticeably. As the refined composite algorithm significantly increases the computation times for these two clinical datasets, the basic versions of multiscale methods are preferable in this case.

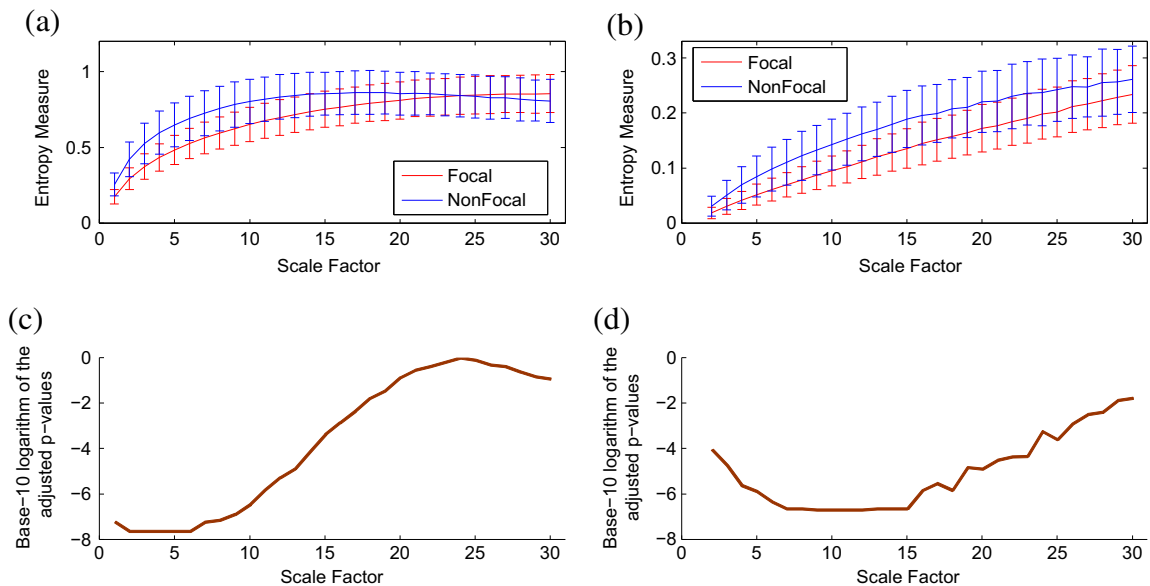

**Fig. 9** Plots illustrating the mean ± SD (with error bars) of the **a** RCMFE$_\mu$ and **b** RCMFE$_\sigma$ values computed from focal and non-focal EEG signals. Base-10 logarithm of the FDR-adjusted $p$ values for the differences in **c** RCMFE$_\mu$ and **d** RCMFE$_\sigma$ at each temporal scale between focal and non-focal signals

**Table 3** The FDR-adjusted $p$ values for focal versus non-focal EEG signals and AD patients' versus controls' MEG recordings of the proposed and classical multiscale entropy-based analyses at scale factor 10

| Dataset | $MSE_\mu$ | $MFE_\mu$ | $RCMSE_\mu$ | $RCMFE_\mu$ |
|---|---|---|---|---|
| EEG data | $2.33 \cdot 10^{-8}$ | $1.18 \cdot 10^{-8}$ | $2.21 \cdot 10^{-8}$ | $1.10 \cdot 10^{-8}$ |
| MEG data | 0.6601 | 0.1321 | 0.6464 | 0.1208 |
| | $MSE_\sigma$ | $MFE_\sigma$ | $RCMSE_\sigma$ | $RCMFE_\sigma$ |
| EEG data | $2.33 \cdot 10^{-10}$ | $1.70 \cdot 10^{-10}$ | $1.99 \cdot 10^{-10}$ | $1.38 \cdot 10^{-10}$ |
| MEG data | 0.0037 | 0.0035 | 0.0036 | 0.0028 |
| | $MSE_\sigma^2$ | $MFE_\sigma^2$ | $RCMSE_\sigma^2$ | $RCMFE_\sigma^2$ |
| EEG data | $2.24 \cdot 10^{-9}$ | $1.07 \cdot 10^{-9}$ | $1.39 \cdot 10^{-9}$ | $1.77 \cdot 10^{-9}$ |
| MEG data | 0.0045 | 0.0052 | 0.0045 | 0.0051 |

## 4 Discussions

In this section, we discuss the results obtained by the existing and proposed multiscale methods for noise and synthetic signals and clinical datasets.

### 4.1 Noise signals

The patterns for $MFE_\sigma$ and $MFE_\sigma^2$ are similar to $MSE_\sigma$ and $MSE_\sigma^2$, respectively. However, as expected theoretically, the SD of $MFE_\sigma$ and $MFE_\sigma^2$ values for each scale is comparatively smaller than that of $MSE_\sigma$ and $MSE_\sigma^2$ measures, respectively. The FuzEn and SampEn for $1/f$ noise are larger than those of WGN when $33 < \tau$ and $42 < \tau$, respectively. It shows another relative advantage of $MFE_\sigma$ over $MSE_\sigma$. Although the curves for $RCMSE_\sigma$ and $RCMSE_\sigma^2$ have smaller SDs than $MSE_\sigma$ and $MSE_\sigma^2$, respectively, for each scale factor, these have larger SDs in comparison with $RCMFE_\sigma$ and $RCMFE_\sigma^2$. This fact confirms our theoretical expectation about $RCMFE_\sigma$ and $RCMFE_\mu$ producing the most stable results among these 12 multiscale entropy methods. In brief, FuzEn-based multiscale methods are more stable than SampEn-based algorithms. Furthermore, the refined composite coarse-graining techniques improve the stability of MSE or MFE. In addition, for $1/f$ noise and WGN time series, the multiscale methods based on standard deviation may have better performance in shorter temporal scales than those based on variance.

### 4.2 Sensitivity of multiscale methods to signal length

Using the fuzzy membership function and/or refined composite technique causes the $RCMFE_\mu$ to become more reliable and stable for short signals in comparison with the other mean-based multiscale methods. Note that the results obtained by variance- and standard deviation-complexity measures are similar to Figs. 3, 4, 5, and 6, although like Fig. 2, the crossing points are different. That is, we have similar advantages of RCMFE methods based on standard deviation or variance over their MSE, MFE, and RCMSE counterparts.

### 4.3 Synthetic signals

For the chirp signal with constant amplitude, the refined composite multiscale entropy-based approaches, i.e., $RCMFE_\sigma$ and $RCMFE_\mu$, are more stable than their basic counterparts ($MSE_\sigma^2$ and $MSE_\mu$). For the abovementioned $AR(1)$ process, results obtained by $MFE_\sigma$ and $MFE_\sigma^2$ have similar patterns, although $MFE_\sigma^2$ is relatively more variable than $MFE_\sigma$. As expected theoretically, refined composite technique reduces the variability of the results. For the aforementioned MIX process, when the TW moves from a stochastic signal to periodic deterministic sequence, the entropy measures for all these methods decrease. In addition, moving from $\tau = 2$ to $\tau = 15$, the entropy measures decrease. Although all these approaches generally demonstrate the same behavior, the $RCMFE_\sigma$ and $RCMFE_\mu$ results are more stable than their corresponding basic counterparts.

For the abovementioned logistic map, the results again demonstrate that mean- and standard deviation-based multiscale approaches, extracting different kinds of dynamical properties of, respectively, mean and spread over multiple time scales, lead to different kinds of features.

The results obtained using the Lorentz system show that although at smaller-scale factors the entropy measures for $RCMFE_\sigma$, $RCMFE_\mu$, $MSE_\mu$, and $RCMSE_\mu$ are very low, two different segments are distinguishable in larger-scale factors. This fact depicts the importance of multiscale entropy methods and temporal scales, in comparison with basic entropy approaches having only scale factor 1, in signal processing. As can be seen in Fig. 7q, t, the results obtained by the $RCMFE_\mu$ are more stable than $RCMSE_\mu$, and $RCMSE_\mu$ results are more stable than $MSE_\mu$ ones. It demonstrates the importance of fuzzy entropy and refined composite algorithm to improve the stability of the results.

### 4.4 Clinical datasets

For MEG dataset, the adjusted $p$ values illustrate that the most significant differences are seen around temporal scales 7–14 and 3–9 using $RCMFE_\mu$ and $RCMFE_\sigma$, respectively. It shows that if a mean-based multiscale entropy cannot discriminate two groups at specific scale factors, its corresponding standard deviation-based one may be able to do so, and vice versa.

The profiles in Fig. 8 show increases in entropy for $RCMFE_\mu$ and $RCMFE_\sigma$ at scales 1–3 and 2–6, respectively. This emphasizes the suitability of multiscale evaluations for the assessment of biomedical data as these approaches managed to reveal different dynamics associated with pathology (AD in this case), despite previous claims by some authors that the coarse-graining procedure in MSE had the shortcoming that it tended to artificially decrease the entropy values as a function of the time scale [39].

For focal and non-focal EEG dataset, since all EEG signals were band-pass filtered between 0.5 and 40 Hz, there is no relevant information left for analysis of frequencies higher than 40 Hz. However, the frequency that corresponds to the analysis of scale 1 is 512/2 Hz. This may be the reason why SampEn is so low for short time scales.

It should be added that the entropy parameters used for biomedical signals are exactly similar to those mentioned for synthetic time series. We tested different $r$ values from 0.05 to 0.2 for this kind of signals, and for all of them, the results had similar patterns and the conclusions do not change when the parameters are varied.

Note that in the MSE algorithm, we kept the value of $r$ fixed across temporal scales. Other authors suggested recalculating the tolerance $r$ at each scale factor separately [14]. Using several physiological datasets, they found that recalculating $r$ produced similar results to those obtained by not recomputing $r$, as in the original description of MSE proposed by Costa et al. [9]. Considering that there was no evidence of the fact that recomputing $r$ for each scale improved the results, we decided to keep $r$ fixed so that we retained the advantages of the original formulation of multiscale entropy by which the entropy of WGN decreases with $\tau$. (Note that renormalizing $r$ for each scale will lead to flatter MSE curves for WGN, contrary to theoretical expectations.)

## 5 Conclusions

In this paper, we introduced the $RCMFE_\sigma$ and $RCMFE_\mu$, extracting different kinds of dynamical properties (or features) of spread and mean, respectively, over multiple time scales. We illustrated the behavior of these multiscale entropy-based approaches versus WGN, $1/f$ noise, several straightforward concepts in signal processing, and two clinical datasets. The results showed that $MSE_\sigma$ and $MFE_\sigma$ had better performance to show the concept of complexity than, respectively, $MSE_\sigma^2$ and $MFE_\sigma^2$ for $1/f$ noise and WGN time series. The FuzEn-based multiscale methods were more stable than SampEn-based algorithms, and furthermore, the refined composite technique noticeably improved the stability of the basic MSE and MFE methods. The proposed methods alleviated the problem of undefined MSE and RCMSE values for short signals. The classification results, obtained using simple classification methods, showed that $RCMFE_\sigma$-based features lead to higher classification accuracies in comparison with the $RCMFE_\mu$-based ones. The results also illustrated that when the $(RC)MFE_\mu$, as a signal-dependent method, cannot distinguish different types of dynamics of a particular signal, the $(RC)MFE_\sigma$ may do so, and vice versa. We expect that our developments will find applications in physiologic and non-physiologic studies to distinguish different kinds of dynamics.

**Acknowledgements** The authors would like to thank Dr. Madalena Costa from the Beth Israel Deaconess Medical Center, Harvard Medical School, Boston, USA, for very useful suggestions and comments. They extend thanks to Dr. Andrzejak from the Department of Information and Communication Technologies, Universitat Pompeu Fabra, Barcelona, Spain, for providing the EEG data described in Sect. 2.

**Compliance with ethical standards** All control subjects and AD patients' caregivers gave informed consent for participation in the study, which was approved by the local Ethics Committee. Retrospective EEG data analysis has been approved by the ethics committee of the Kanton of Bern. Moreover, all patients gave written informed consent that the obtained signals from long-term EEG might be utilized for research purposes.

## Appendix

The codes for our analysis, including SampEn, FuzEn, $MSE_\mu$, $MFE_\mu$, $RCMSE_\mu$, $RCMFE_\mu$, $MSE_\sigma^2$, $MFE_\sigma^2$, $RCMSE_\sigma^2$, $RCMFE_\sigma^2$, $MSE_\sigma$, $MFE_\sigma$, $RCMSE_\sigma$, and $RCMFE_\sigma$, are available at http://dx.doi.org/10.7488/ds/1477.

**Hamed Azami** is a PhD student at The University of Edinburgh, UK. His research interest is developing signal processing and pattern recognition techniques with major applications in biomedical data and neuroscience.

**Alberto Fernández** is "Profesor Titular" in the Department of Psychiatry, Complutense University of Madrid. His research interests include the application of magnetoencephalography to the diagnosis of Alzheimer's disease.

**Javier Escudero** is a tenured faculty member (Chancellor's Fellow) at the University of Edinburgh, UK. His research interests include biomedical signal processing and pattern recognition in clinical applications.